\documentclass[aps,prl,twocolumn,showpacs,preprintnumbers,superscriptaddress]{revtex4}

\usepackage{graphicx}% Include figure files
\usepackage{epsf} % Include figure files, old times style !!
\usepackage{dcolumn}% Align table columns on decimal point
\usepackage{bm}% bold math
\usepackage{amssymb}
\usepackage{amsmath}
\usepackage{times}

\begin{document}

\title{Universal Critical Behavior of the Synchronization
Transition in Delayed Chaotic Systems}

\author{Ivan G.\ Szendro} \email{szendro@ifca.unican.es}

\affiliation{Instituto de F\'{\i}sica de Cantabria (IFCA),
CSIC--UC, E-39005 Santander, Spain}

\affiliation{Departamento de F{\'\i}sica Moderna, Universidad de
Cantabria, Avda. Los Castros, E-39005 Santander, Spain}

\author{Juan M.\ L{\'o}pez} \email{lopez@ifca.unican.es}

\affiliation{Instituto de F\'{\i}sica de Cantabria (IFCA),
CSIC--UC, E-39005 Santander, Spain}

\date{\today}

\begin{abstract}
We numerically investigate the critical behavior of the
synchronization transition of two unidirectionally coupled delayed
chaotic systems. We map the problem to a spatially extended system
to show that the synchronization transition in delayed systems
exhibits universal critical properties. We find that the
synchronization transition is absorbing and generically belongs to
the universality class of the bounded Kardar-Parisi-Zhang
equation, as occurs in the case of extended systems. We also argue
that directed percolation critical behavior may emerge for systems
with strong nonlinearities
\end{abstract}


\maketitle Synchronization of chaotic systems has attracted much interest in recent years
and examples include chemical reactions, neuronal networks, Josephson junctions,
electronic circuits, and semiconductor lasers, among others (see Ref.\cite{syn-review}
and references therein). More recently, and from a practical point of view, this burst of
activity in the field is partially due to the potential applications in control and
secure communications. It is expected that an increased complexity of the attractor would
make it much more difficult to extract the dynamical information. In particular, delayed
dynamical systems have been suggested as the ideal candidates for secure communication
for several reasons. On the one hand, they are hyperchaotic systems with an arbitrarily
large number of positive Lyapunov exponents, whose number increases linearly with the
delay time \cite{farmer,bunner}. On the other hand, they may be experimentally realized
in the form of fast communication optical systems by using different types of delayed
feedback setups \cite{ikeda,roy,claudio,abarbanel,goedgebuer1,goedgebuer2,jordi}.

Synchronization of two separate delayed chaotic systems is
achieved by allowing some communication between them. The possible
schemes are diverse, including variable substitution, symmetric
feedback, {\it etc}. Generally, there exists a critical coupling
constant $\kappa_c$ that separates two different phases. For low
coupling values, $\kappa < \kappa_c$, there is a {\em disordered
phase} in which each system evolves independently and the time
average difference between both systems remains finite. In
contrast, a {\em synchronized phase} appears for $\kappa >
\kappa_c$, in which the average error tends to zero and memory of
the initial difference is asymptotically lost.

Very recent studies have been devoted to investigate important
aspects of synchronization in delayed dynamical systems as, for
instance, analytical approximations to estimate the
synchronization threshold \cite{pyragas}, the robustness of the
transition to parameter mismatch \cite{bocca}, chaos control in
lasers with feedback \cite{blakely}, information flow between
drive and response systems \cite{pethel}, and the effect of a
time-dependent delay \cite{kye}. However, the mechanism behind the
synchronization transition in delayed dynamical systems and its
relationship with those exhibited by other chaotic systems with
many degrees of freedom is still unknown.

Remarkably, synchronization also takes place in coupled spatially
extended systems with many degrees of freedom and space-time chaos.
In this case, the synchronization transition has been shown to be an
{\em absorbing} nonequilibrium phase transition and, accordingly,
its critical properties have attracted much interest in the past few
years \cite{grassberger,baroni,ahlers,droz,lipo,mamunoz}. Despite
being scalar ({\it i.e.}, described by only one dynamical variable),
delayed dynamical systems are formally infinite dimensional
dynamical systems and show many aspects of space-time chaos,
including the formation and propagation of structures, defects and
spatiotemporal intermittency
\cite{arecchi2,giacomelli1,giacomelli2,sanchez}. An interesting
question that naturally arises is whether the synchronization
transition in (scalar) hyper-chaotic systems with delayed feedback
could also be understood as a nonequilibrium phase transition, as
occurs in (vectorial) extended dynamical systems with space-time
chaos.

In this Letter, we characterize the synchronization transition in
unidirectionally coupled delayed dynamical systems as a
nonequilibrium critical phase transition and relate it to existing
universality classes. We exploit the interpretation of delayed
dynamical systems as spatially extended systems
\cite{arecchi2,giacomelli1,giacomelli2,sanchez} to show that the
synchronization transition in delayed systems exhibits {\em
universal} properties, which are independent of microscopic details
of the individual systems being coupled. We find that the
synchronization transition generically belongs to the universality
class of the bounded Kardar-Parisi-Zhang (BKPZ) equation, as occurs
in the case of extended systems
\cite{grassberger,baroni,ahlers,droz,lipo,mamunoz}. We also argue
that directed percolation (DP) critical behavior may emerge for
systems with strong nonlinearities. Our results show that the
critical properties of the synchronization transition in delayed
chaotic systems are identical to those in spatially extended
systems, despite being the former a scalar system with no real
spatial structure.

We consider two identical time-delay systems described
by two coupled differential-delay equations, the drive
(transmitter) system
\begin{equation}
\label{dds1} \dot{u}= -a\,u + F (u_\tau),
\end{equation}
and the response (receiver) system
\begin{equation}
\label{dds2} \dot{v}= -a\,v + F (v_\tau) + \kappa (u-v),
\end{equation}
where $u_\tau=u(t-\tau)$ and $v_\tau=v(t-\tau)$ are the delayed
variables, $\tau$ is the time delay, and $\kappa$ is the coupling
strength.

Delayed systems like Eq.(\ref{dds1}) are used in a variety of
applications ranging from biology to optics. We have studied in
detail three prototypical models: the Mackey-Glass model \cite{mg},
$F(\rho) = b\, \rho/(1+\rho^{10})$, (initially introduced to
describe regulation of blood cell production in patients with
leukemia), the Ikeda differential-delay equation \cite{ikeda},
$F(\rho) = b\, \sin(\rho-\rho_0)$ (which appears in the context of
optical feedback on a laser-beam \cite{ikeda} and experimental
setups of optical generators of chaos in wavelength
\cite{goedgebuer1,goedgebuer2}) and a model with the piecewise
linear delay-expression given by $F(\rho) = 2 \rho $ if $\rho \leq
1/2$ and $F(\rho) = 2 - 2 \rho$ if $\rho > 1/2$. We have studied the
synchronization critical properties by means of computer simulations
of these three systems and found similar results.

\begin{figure}
\centerline{\epsfxsize=8.5cm \epsfbox{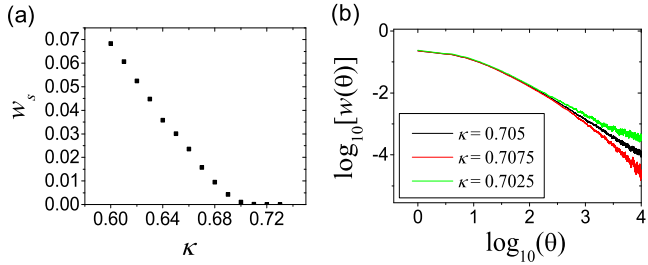}} \caption{(Color online) In (a) the
average synchronization error in the stationary state, $w_s$, is plotted near the
synchronization threshold for $\tau =200$. Each point corresponds to an average over 600
realizations. (b) shows the spatial average of the synchronization error, $w ( \theta )$,
for $\tau =2000$ and $3$ different values of $\kappa$, each curve being an average over
300 realizations. $\kappa_c = 0.705$ is obtained and the corresponding slope yields
$\delta = 1.14$. }
\end{figure}

In order to make apparent the existence of a nonequilibrium phase
transition we transform the pair of coupled delayed systems
Eqs.(\ref{dds1}) and (\ref{dds2}) into two coupled spatially
extended systems. This can be readily done by introducing the
coordinate transformation, $t=x+\theta\,\tau$, where $x\in [0,\tau]$
is the {\em space variable}, while $\theta\in \mathbb{N}$ is a
discrete time variable \cite{arecchi2}. Note that the time delay
$\tau$ becomes the {\em system size}, in such a way that the time
dependence with the delayed variable is transformed into an
interaction within the horizontal space coordinate $x$ in the
space-time representation. This is a powerful representation in
which delayed systems can be treated as extended systems to identify
many features of space-time chaos
\cite{arecchi2,giacomelli1,giacomelli2}.

Complete synchronization of the two coupled delayed systems,
Eqs.(\ref{dds1}) and (\ref{dds2}), is achieved if the
synchronization error $u(t) - v(t)$ vanishes for all times $t$ as
$t \to \infty$. In the spatial picture we replace the dynamical
variables $u(t)$ and $v(t)$ by $\tilde{u}(x,\theta)$ and
$\tilde{v}(x,\theta)$, so that the synchronization error is given
by $w(x,\theta) = \tilde{u}(x,\theta) - \tilde{v}(x,\theta)$, and
synchronization occurs when $w(x,\theta)$ vanishes at all $x$ for
$\theta \to \infty$. This is equivalent to a vanishing spatial
average $\langle \vert w(x,\theta) \vert \rangle_x$. Note that the
spatial average in the space-time representation corresponds to
the average within the delay time $\tau$. In contrast, one has
$\langle \vert w(x,\theta) \vert \rangle \rangle_x > 0$ in the
unsynchronized state. This makes the average error $w(\theta)
\equiv \langle \vert w(x,\theta) \vert \rangle_x$ a natural order
parameter for the transition. Critical properties of the
synchronization transition can now be studied by analyzing the
dependence of the order parameter on the coupling strength
$\kappa$. In addition, the analysis of the critical behavior for
finite time delays can be naturally carried out by standard
finite-size scaling techniques. The remaining part of this Letter
is devoted to the study of these issues.

Our findings are based upon extensive numerical simulations of the
three time-delay systems introduced above. In all our numerical
simulations we have used the Adams-Bashforth-Moulton
predictor-corrector scheme \cite{nr} to integrate the coupled
differential-delay equations (\ref{dds1}) and (\ref{dds2}). For the
sake of brevity we focus the discussion of our numerical results on
the Mackey-Glass model, but we found similar results for the Ikeda
equation and the piecewise linear system. The parameters $a=1$ and
$b=2$ are used in all the results we are presenting here, and
simulation with a time delay varying from ten to a few thousand time
units have been carried out using an integration step of $\Delta t =
0.01$. The region of interest here corresponds to delays $\tau \gg
1.7$ for which the Mackey-Glass model is hyper-chaotic
\cite{farmer}.
%Simulations are computationally very time consuming, since they
%involve numerical integration of differential equations, which
%contrasts with the faster algorithms that can be used to study
%extended systems modelled by coupled-map lattices (which are
%discrete time and discrete space models). This fact severely
%limits the system sizes ({\it i.e.}, time delays) and the
%statistical errors of our simulations as compared with
%coupled-map
%lattices. %Simulations were carried out on a cluster of 20 IBM
%servers (1.26 GHz) and took about 336 hours of computing time.

\begin{figure}
\centerline{\epsfxsize=8.5cm \epsfbox{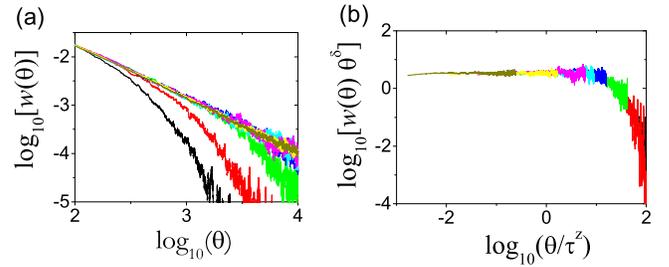}} \caption{(Color online) Finite-size data
in (a) unscaled and (b) scaled coordinates for $\tau \in  \{ 10, 20, 50, 100, 200, 150,
500, 2000 \} $ and $\kappa_c = 0.705$. The collapse is obtained for $\delta = 1.15$ and
$z = 1.45$. Every curve corresponds to an average over 60 realizations.}
\end{figure}

In Fig.\ 1a we present our results for the order parameter, {\it
i.e.}, the average synchronization error, in the stationary state
$w_s (\kappa) = \lim_{\theta\to\infty}\langle \vert w (x,\theta)
\vert \rangle_{x,\theta}$ for a system of size (delay) $\tau =
200$ as the coupling strength is varied. Inspection of Fig.\ 1a
indicates that the transition is continuous and occurs roughly
around $\kappa = 0.7$, which is in agreement with earlier
estimations for the same model parameters \cite{pyragas}. Dynamic
critical behavior is studied by calculating the indices $\delta$
and $\beta$ that describe the critical behavior of the order
parameter near the threshold for synchronization: $w(\theta) \sim
\theta^{-\delta}$ for $\kappa = \kappa_c$ and $w_s(\kappa) \sim
\vert \kappa - \kappa_c \vert^\beta$ as the transition is
approached for $\kappa < \kappa_c$. Studying critical behavior
demands to obtain a good estimation for the critical threshold
which implies the use of large delays (system sizes). In Fig.\ 1b
we plot the sub-critical and super-critical behavior of the
average synchronization error for a large system size $\tau =
2000$ as the transition is approached from below ($\kappa =
0.7025$) and above ($\kappa = 0.7075$), respectively. Within our
numerical resolution we find that the best power-law behavior
$w(\theta) \sim \theta^{-\delta}$ is obtained at $\kappa_c = 0.705
\pm 0.002$, which gives an estimation for the critical exponent
$\delta = 1.14 \pm 0.03$.

\begin{figure}
\centerline{\epsfxsize=8.5cm \epsfbox{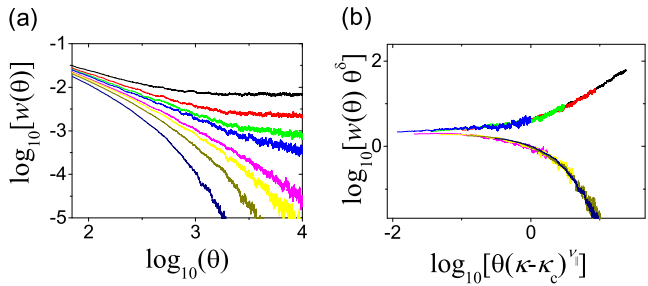}} \caption{(Color online) Off-critical
data in (a) unscaled and (b) scaled coordinates for $\tau = 2000$ and various values of
$\kappa \in \lbrack 0.685, 0.725 \rbrack$. The collapse is obtained for $\kappa_c =
0.705$, $\delta = 1.05$ and $\nu_{\parallel} =1.4$. Every curve corresponds to an average
over 60 realizations.}
\end{figure}

Once the critical threshold has been determined, we use
finite-size scaling at the critical point $\kappa_c = 0.705$ and
fit numerical data to the scaling form
\begin{equation}
w(\theta) = \theta^{-\delta} f(\theta/\tau^z), \label{fss}
\end{equation}
where the scaling function $f(y) \sim {\rm const}$ for $y \ll 1$
and $f(y) \sim y^{\delta}$ for $y \gg 1$. This gives us an
independent determination of $\delta$ and the dynamic exponent
$z$. In Fig. 2a we show numerical results for different system
sizes at the critical point $\kappa_c$ and these data are best
collapsed in Fig. 2b with $z=1.45 \pm 0.05$ and $\delta = 1.15 \pm
0.05$ (the latter in good agreement with our previous estimate in
Fig.\ 1).

Next we report on off-critical numerical calculations of the
synchronization error. This allows us to estimate the correlation
length exponent. In Fig.\ 3a we plot the order parameter as the
coupling strength is varied close to the synchronization threshold
for a large system size $\tau = 2000$. Near and below the
transition the characteristic size of synchronized regions within
the disordered phase is given by the horizontal correlation length
$\xi$ and is expected to diverge as $\xi \sim
\epsilon^{-\nu_{\perp}}$ when the distance to the critical point
tends to zero, $\epsilon = \vert \kappa - \kappa_c \vert \to 0$.
Correspondingly, the characteristic time $\vartheta$ measuring the
duration of a fluctuation of size $\xi$ diverges as $\vartheta
\sim \xi^z \sim \epsilon^{-\nu_{\parallel}}$, where
$\nu_{\parallel} = \nu_{\perp} z$. Off-critical data are then
expected to satisfy the scaling form $w(\theta,\epsilon) =
\theta^{-\delta} g(\theta/\vartheta)$, so that numerical data in
Fig.\ 3a must collapse according to
\begin{equation}
w(\theta,\epsilon) = \theta^{-\delta} g(\theta
\epsilon^{\nu_{\parallel}}) \label{off-crit}
\end{equation}
for the appropriate election of the critical exponents $\delta$
and $\nu_{\parallel}$. In Fig.\ 3b we show a data collapse for the
exponents $\delta = 1.05 \pm 0.05$ and $\nu_{\parallel} = 1.4 \pm
0.1 $ where the two branches correspond to numerical data for
coupling strengths above and below critical. Also the index
$\beta$ can be obtained from the scaling behavior Eq.\
(\ref{off-crit}), for $\theta \gg \vartheta$ we have $w(\theta \to
\infty, \epsilon) \sim \epsilon^{\beta}$, where $\beta =
\nu_{\parallel}\delta = 1.47$.

\begin{figure}
\centerline{\epsfxsize=7.0cm \epsfbox{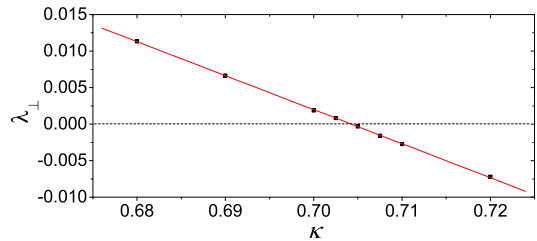}} \caption{(Color online) The transverse
Lyapunov exponent, $\lambda_{\perp}$, is plotted for $\tau = 2000$ and various values of
$\kappa$ around the synchronization threshold. Note that $\lambda_{\perp}$ changes sign
near $\kappa_c$.}
\end{figure}

The critical exponents of the synchronization transition in
delayed systems are to be compared with those observed in extended
systems. This will allow us to identify the mechanisms behind the
transition in both types of high-dimensional dynamical systems. In
the context of extended systems, the exponential growth rate of
the error $\vert w (x,t) \vert$ is known as the transverse
Lyapunov exponent $\lambda_{\perp}$ and measures the stability of
the synchronized solution $w(x,t) = 0$. Accordingly, stable
synchronization implies that the transverse Lyapunov exponent must
be negative. For dynamical systems with {\em smooth} local
nonlinearities (like lattices of logistic or tent coupled maps),
this proves to be a sufficient condition as well. In this case,
the synchronization transition is found to be generically in the
universality class of the KPZ equation with a (bounding)
growth-limiting term, the so-called BKPZ universality
\cite{baroni,ahlers,droz,lipo,mamunoz}. Numerical estimates of the
critical exponents gave $\delta_{BKPZ} = 1.17 \pm 0.05$,
$\beta_{BKPZ} = 1.50 \pm 0.05$ and $z_{BKPZ} = 1.53 \pm 0.05$ for
different models studied in the recent literature
\cite{ahlers,droz,lipo,mamunoz}. On the contrary, in the presence
of {\em strong} and localized nonlinearities (like for instance
for Bernoulli coupled maps), the synchronized phase turns out to
be unstable even for negative values of $\lambda_{\perp}$
\cite{ahlers}. In this case, the transition occurs only when the
propagation velocity of finite-amplitude perturbations vanishes.
The critical properties of the transition are then associated with
the DP universality class \cite{ahlers,mamunoz}. The fraction of
non-synchronized sites corresponds to the fraction of active sites
in DP. Correspondingly, the critical exponents are given by
$\delta_{DP} = 0.159$, $\beta_{DP} = 0.277$, $z_{DP} = 1.581$
\cite{ahlers,mamunoz}. The DP correlation length and time
exponents are known to be $\nu_{\perp} = 1.10$ and
$\nu_{\parallel} = 1.73$, respectively \cite{dp}.

Our numerical results, $\delta= 1.15$, $\beta = 1.47$, $z= 1.45$
and $\nu_{\parallel} = 1.14$, strongly suggest that the
synchronization transition in delayed chaotic systems generically
belongs to the BKPZ universality class, as occurs in extended
chaotic systems. As an additional check, we have measured the
transverse Lyapunov exponent $\lambda_{\perp}$ for the coupled
delayed systems, Eqs.\ (\ref{dds1}) and (\ref{dds2}), with
different coupling strengths and, as showed in Fig.\ 4, we found
that the transition takes place when $\lambda_{\perp}$ becomes
negative, as expected for the BKPZ universality class.
Nevertheless, the nature of the transition can be changed to DP
behavior in the presence of strong local nonlinearities akin to
what occurs in extended systems. In fact, by choosing the
nonlinear function $F(\rho) = 2 \rho \mod{1}$ the exponent
$\delta$ drops to $\delta = 0.16 \pm 0.03$ in good agreement with
DP. The different nature of the transitions in the two cases can
be seen in Fig.\ 5, where we show the spatiotemporal evolution of
the synchronization error $\vert w(x,\theta) \vert$ for the (a)
smooth Mackey-Glass and (b) strongly nonlinear model,
respectively, just slightly above the transition.

\begin{figure}
\centerline{\epsfxsize=8.5cm \epsfbox{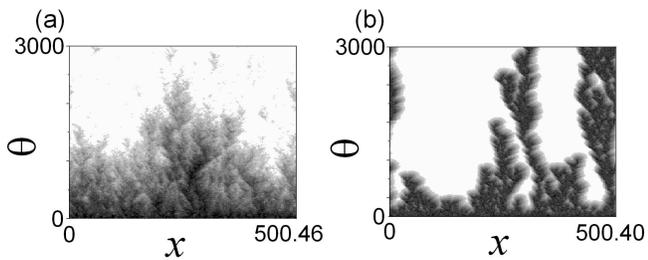}} \caption{The synchronization error for a
coupling slightly above the synchronization threshold, $\kappa \gtrsim \kappa_c$, is
plotted for $\tau = 500$ in the cases of (a) smooth nonlinearities and (b) strong local
nonlinearities. The horizontal width has been chosen slightly larger than $\tau$ to
eliminate the systematic drift. Compare with Fig.2 of Ref.\cite{ahlers}}
\end{figure}

In conclusion, we have studied the critical properties of the
synchronization transition in unidirectionally coupled delayed
chaotic systems. We used a standard coordinate transformation to map
the (scalar) time-delay system to a spatially extended system. This
mapping allowed us to study the synchronization transition as a {\em
absorbing} nonequilibrium phase transition. Comparison of the
critical exponents as well as the behavior of the transverse
Lyapunov exponent lead us to conclude that the synchronization
transition in delayed systems generically belongs to the BKPZ
universality class independently of the specific form of the
delay-expression, as long as it is a continuous function, just as
occurs for synchronization of space-time chaos. Finally, our
numerical results also indicate that the existence of
discontinuities in the delay nonlinear function may change this
critical behavior from BKPZ to DP, which suggests that the same
mechanisms that produce DP behavior in coupled extended systems can
be invoked in the case of delayed chaotic systems.

\begin{acknowledgements}
The authors acknowledge computing time on the "Santander GridWall"
cluster at IFCA-CSIC. This work is supported by the MCyT (Spain)
through Grant No. BFM2003-07749-C05-03 and the European Commission
through Grant No. OCCULT IST-2000-29683. I.\ S.\ thanks additional
support from the Vicerrectorado de Investigaci\'on (Universidad de
Cantabria).

\end{acknowledgements}

\end{document}